\documentclass[pra,twocolumn,amsmath,amssymb]{revtex4}

\usepackage{graphicx,epsfig}

\newcommand{\ket}[1]{|#1\rangle} 

\def\be{\begin{equation}}
\def\ee{\end{equation}}
\def\bea{\begin{eqnarray}}
\def\eea{\end{eqnarray}}
\def\bma{\begin{mathletters}}
\def\ema{\end{mathletters}}

\def\0{\overline{0}}

\def\q0{\underline{0}}

\def\one{\leavevmode\hbox{\small1\normalsize\kern-.33em1}}
\def\compl{\begin{picture}(8,8)\put(0,0){C}\put(3,0.3){\line(0,1){7}}\end{picture}}

 \def\ket#1{|#1\rangle}

\begin{document}

\title{Entanglement Percolation in Quantum Networks}
\author{Antonio Ac\'\i n}
\affiliation{ICFO--Institut de Ciencies Fotoniques, E-08860
Castelldefels, Barcelona, Spain}
\author{Ignacio Cirac}\affiliation{Max-Planck-Institut f\"ur Quantenoptik, Garching}
\author{Maciej Lewenstein}
\affiliation{ICREA and ICFO--Institut de Ciencies Fotoniques,
E-08860 Castelldefels, Barcelona, Spain}

\begin{abstract}
Quantum networks are composed of nodes which can send and receive quantum states by exchanging photons \cite{Mabuchi}. Their goal is to facilitate quantum communication between any nodes, something which can be
used to send secret messages in a secure way
\cite{BB84,Ekert}, and to communicate more efficiently than
in classical networks \cite{finger}. These
goals can be achieved, for instance, via teleportation \cite{Wootters}.  Here we show that the design of efficient quantum communication protocols in quantum networks involves intriguing quantum phenomena,
depending both on the way the nodes are displayed,  and the
entanglement between them. These phenomena can be employed  to design protocols which overcome the exponential decrease of signals with the number
of nodes. We relate the problem of establishing
maximally entangled states between nodes to classical
percolation in statistical mechanics \cite{Grimmett},
and demonstrate that quantum phase transitions \cite{Sachdev99}
can be used to optimize the operation of quantum networks.
\end{abstract}


\maketitle

\noindent{\bf Introduction.} The future of quantum communication will be based on quantum
networks (cf. \cite{Mabuchi,Knill,Duan,Briegel,Kuzmich,Gisin,Paivi,Leung}),
where different nodes are entangled, leading to quantum
correlations which can be exploited by performing local
measurements in each node. For instance, a set of quantum
repeaters \cite{Briegel} can be considered as a simple quantum network where the
goal is to establish quantum communication over long distances. In
order to optimize the operation of such a network, it is required
to establish efficient protocols of measurements in such a way
that the probability of success in obtaining maximally entangled
states between different nodes is maximized. This probability may
behave very differently as a function of the number of nodes if we
use different protocols: in some cases it may decay exponentially,
something which makes the repeaters useless, whereas for some
protocols it may decay only polynomially, something which would
make them very efficient.

A general network may be characterized by a quantum state, $\rho$,
shared by the different nodes. The goal is then: given two nodes,
$A$ and $B$, find the measurements to be performed in the nodes,
assisted with classical communication, such that $A$ and $B$ share
a maximally entangled state, or singlet, with maximal probability.
We call this probability the singlet conversion probability (SCP).
This, or other related quantities like the localizable
entanglement \cite{LE,LE2}, can be used as a figure of merit to
characterize the state $\rho$ and therefore the performance of the
quantum network. Here, we focus on the SCP because of its
operational meaning. These quantities cannot be determined in
general, given that they require the optimization over all
possible measurements in the different nodes, which is a
formidable task even for small networks.

In this work we concentrate in some particular quantum networks
which, despite its apparent simplicity, contain a very rich and
intriguing behavior. The simplification comes from two facts (see
Fig. 1): first, the nodes are spatially distributed in a regular
way according to some geometry. Second, each pair of nodes are
connected by a pure state
$\ket{\varphi}\in\compl^d\otimes\compl^d$. Up to local change of
bases, any of these states can be written as \cite{NC}
\begin{equation}\label{stated}
    \ket{\varphi}=\sum_{i=1}^d\sqrt{\lambda_i}\ket{ii} ,
\end{equation}
where $\lambda_i$ are the (real) Schmidt coefficients such that
$\lambda_1\geq\lambda_2\geq\ldots\geq\lambda_d\geq 0$.  This
configuration reminds of the states underlying the so-called
projected pair entangled states \cite{PEPS}, and thus we call
these networks pair-entangled pure networks (PEPN). For these
geometries, we first introduce a series of protocols which are
closely related to classical percolation  \cite{Grimmett}, a
concept that appears in statistical mechanics. We then determine
the optimal protocols for several 1-dimensional (1D)
configurations, where some counterintuitive phenomena occur. We
use these phenomena to introduce various protocols in more complex
2-dimensional (2D) configurations.

We show that these new protocols provide a dramatic improvement
over those based on classical percolation, in the sense that one
can obtain perfect quantum communication even though the
percolation protocols give rise to an exponential decay of the
success probability with the number of nodes. In fact, we will
argue that there exists a quantum phase transition in the quantum
networks which may be exploited to obtain very efficient
protocols. Thus, this work opens a new set of problems in quantum
information theory which are related to statistical physics, but
pose completely new challenges in those fields. As opposed to most
of the recent work on entanglement theory, which has been devoted
to using some of the tools developed so far in quantum information
theory to analyze problems in statistical mechanics \cite{PEPS,
Fazio,Osborne,VidalLatorre}, the present work makes a step in the
converse direction.

\noindent{\bf Classical entanglement percolation.} A first natural
measurement consists of all the pairs of nodes nodes locally
transforming their states into singlets with optimal probability,
$p\,^{ok}$. Recall that the SCP for a state (\ref{stated}) is
known to be equal to $p\,^{ok}=\min(1,2(1-\lambda_1))$
\cite{Vidal}. Then, a perfect quantum channel between the nodes is
established with probability $p\,^{ok}$ , otherwise no
entanglement is left. This problem is equivalent to a standard
bond percolation situation \cite{Grimmett}, where one distributes
connections among the nodes of a lattice in a probabilistic way:
with probability $p$ an edge connecting a pair of nodes is
established, otherwise the nodes are kept unconnected. We call
this measurement strategy classical entanglement percolation
(CEP). In bond percolation, for each lattice geometry there exists
a percolation threshold probability, $p_{th}$, such that an
infinite connected cluster  can be established if and only if
$p>p_{th}$ (see also Table 1). The probability $\theta(p)$ that a
given node belongs to an infinite cluster, or {\it percolation
probability}, is strictly positive for $p\ge p_{th}$,  and zero
otherwise (in the limit of an infinite number of nodes). Then, the
probability that two given distant nodes can be connected by a
path is distance independent and given, correspondingly, by
$\theta^2(p)$ for $p>p_{th}$; for $p<p_{th}$ this probability
decays exponentially with the number of nodes, $N$.

The threshold probabilities define a minimal amount of
entanglement for the initial state such that CEP is possible. In
the case of 1D chains, see Fig. \ref{repeater}, percolation is
possible if, and only if, $p=1$. Therefore, the SCP decays
exponentially with $N$ unless the states are more entangled than
the singlet, in the sense that $p\,^{ok}=1$. In a square 2D lattice, the
entanglement threshold derived from percolation arguments, see
Table \ref{percpr}, is $p\,^{ok}=2(1-\lambda_1)=1/2$.

It is natural to wonder whether CEP is optimal for any geometry
and number of nodes. And if not, to see if, at least, it predicts
the correct decay of entanglement in the asymptotic limit. Next,
we show that, for 1D chains, although CEP is not optimal for some
finite $N$, it gives the right asymptotic behavior. Moving to 2D
networks, we prove that CEP is not optimal even in the asymptotic
case. Thus, the problem of entanglement distribution through
quantum networks defines a new type of critical phenomenon, that
we call entanglement percolation.

\noindent{\bf 1D Chains}
The scenario of a 1D chain configuration, see Fig. \ref{repeater},
consists of two end nodes connected by several repeaters
\cite{Briegel}. As said, all the bonds are equal to
$\ket{\varphi}$.

We start out with the case of qubits, $d=2$. A surprising result
already appears in the first non-trivial situation consisting of
one repeater. An upper bound to the SCP in the one-repeater
scenario is obtained by putting nodes $A$ and $R_1$ together, which implies that the SCP cannot be larger than $p\,^{ok}$. This bound can indeed be
achieved by means of a rather simple protocol involving
entanglement swapping \cite{Swapping} at the repeater. However, if
CEP is applied, the obtained SCP is simply $(p\,^{ok})^2$. This
proves that CEP is not optimal already for the one-repeater
configuration. We find quite counter-intuitive that the
intermediate repeater does not decreases the optimal SCP. This
behavior, however, does not survive in the asymptotic limit. In
this limit, the so-called concurrence \cite{Concurrence}, another
measure of entanglement, decreases exponentially with the number
of repeaters, unless the connecting states are maximally entangled
(see Methods). The exponential decay of the SCP automatically
follows.

Most of these results can be generalized to higher dimensional
systems, $d>2$. For the one-repeater configuration, the SCP is
again equal to $p\,^{ok}$. It suffices to map the initial state
into a two-qubit state, without changing the SCP, and then apply
the previous protocol. Moving to the asymptotic limit, an
exponential decay of the SCP with $N$ can be proven in the
scenario where the measurement strategies only involve one-way
communication: first, a measurement is performed at the first
repeater. The result is communicated to the second repeater, where
a second measurement is applied. The results of the two
measurements are communicated to the third, and so on until the
last repeater, where the final measurement depends on all the
previous results.

Putting all these results together, a unified picture emerges for
the distribution of entanglement in 1D chains: despite some
remarkable effects for finite $N$, the SCP decreases exponentially
with the number of repeaters whenever the connecting bonds have
less entanglement than a singlet. The CEP strategy fails for some
finite configuration, but predicts the correct behavior in the
asymptotic limit.

\noindent{\bf 2D Lattices.} The situation becomes much richer for
2D geometries. First, we consider finite 2D lattices. The
non-optimality of CEP can be shown already for the simplest
$2\times 2$ square lattice and qubits. Consider the two
non-neighboring sites in the main diagonal of the square. The SCP
obtained by CEP is $1-(1-(p\,^{ok})^2)^2$. By concatenating the
optimal measurement strategy for the one-repeater configuration,
the SCP is $1-(1-p\,^{ok})^2$. However none of these strategies
exploits the richness of the 2D configuration. Indeed, one can
design strategies such that a singlet can be established with
probability one whenever $\ket{\varphi}$ satisfies
$1/2\leq\lambda_1\lessapprox 0.6498$. Thus, there are 2D network
geometries where, although the connections are not maximally
entangled, the entanglement is still sufficient to establish a
perfect quantum channel.

Let us now see whether the thresholds defined by standard
percolation theory are optimal for asymptotically large networks.
In the next lines, we construct an example that goes beyond the
classical percolation picture, proving that the CEP strategy is
not optimal. The key ingredient for this construction is the
measurement derived above for the one-repeater configuration,
which gave raise to a SCP equal to $p\,^{ok}$. Our example
considers a honeycomb lattice where each node is connected by two
copies of the same two-qubit state
$\ket{\varphi}=\ket{\varphi_2}^{\otimes 2}$, see Fig.
\ref{hontr}.a. If, as above, the Schmidt coefficients of the
two-qubit state are $\lambda_1\geq\lambda_2$, the SCP of
$\ket{\varphi}$ is given by $p\,^{ok}=2(1-\lambda_1^2)$. We choose
this conversion probability smaller than the percolation threshold
for the honeycomb lattice, which gives
\begin{equation}\label{l1cr}
    \lambda_1=\sqrt{\frac{1}{2}+\sin\left(\frac{\pi}{18}\right)}\approx 0.82
    .
\end{equation}
So, CEP is useless. Now, half of the nodes perform the optimal
strategy for the one-repeater configuration, mapping the honeycomb
lattice into a triangular lattice, as shown in Fig. \ref{hontr}.
The SCP for the new bonds is exactly the same as for the state
$\ket{\varphi_2}$, that is $2\lambda_2$. This probability is
larger than the percolation threshold for the triangular lattice,
since
\begin{equation}
    2\lambda_2=2\left(1-\sqrt{\frac{1}{2}+\sin\left(\frac{\pi}{18} \right)}\right) \approx 0.358 > 2\sin\left(\frac{\pi}{18}\right) .
\end{equation}
The nodes can now apply CEP to the new lattice and succeed. Thus,
this strategy, which combines entanglement swapping and CEP,
allows to establish a perfect quantum channel in a network where
CEP fails.

\noindent{\bf Conclusions.} We have shown that the distribution of
entanglement through quantum networks defines a framework where
statistical methods and concepts, such as classical percolation
theory and beyond, naturally apply. It leads to a novel type of
quantum phase transitions, that we call entanglement percolation,
where the critical parameter is the minimal amount of entanglement
necessary to establish a perfect distant quantum channel with
significant (non-exponentially decaying) probability. Further
understanding of optimal entanglement percolation strategies is
necessary for the future development and prosperity of quantum
networks.

{\bf METHODS}

\noindent{\bf 1D chains}

We start by showing that the concurrence decays exponentially with
the number of nodes in a 1D chain of qubits when the connecting
states are not maximally entangled. Recall that, given a two-qubit
pure state $\ket{\varphi}=\sum_{i,j}t_{ij}\ket{ij}$, its
concurrence reads $C(\varphi)=2|\det(T)|$, where $T$ is the
$2\times 2$ matrix such that $(T)_{ij}=t_{ij}$.

When considering the repeater configuration, the maximization of
the averaged concurrence turns out to be equal to
\begin{equation}\label{concurrence}
     C_N = \sup_{\cal M}
     \sum_r 2 |\det(\varphi_1 M_{r_1}\varphi_2
     \ldots M_{r_{N}} \varphi_{N+1})|.
\end{equation}
Here ${\cal M}$ briefly denotes the choice of measurements, while
$\varphi_k$ represent the $2\times 2$ diagonal matrices given by
the Schmidt coefficients of the states $\ket{\varphi_k}$.
$M_{r_k}$ are also $2\times 2$ matrices, corresponding to the pure
state $\ket{\mu_k}$ associated to the measurement result $r_k$ of
the $k$-th repeater, that is
$\ket{r_k}=\sum_{i,j}(M_{r_k})_{ij}\ket{ij}$. Note that the
computational basis $i$ and $j$ in the previous expressions are
the Schmidt bases for the states $\ket{\varphi_k}$ and
$\ket{\varphi_{k+1}}$ entering the repeater $k$. Using the fact
that $\det(AB)=\det(A)\det(B)$, the previous maximization gives
\cite{Martin}
\begin{equation}\label{1dqubit}
     C_{N} = \prod_{k=1}^N 2 |\det(\varphi_k)|.
\end{equation}
Note that $|2 \det(\varphi_k)|=1$ if and only if $\ket{\varphi_k}$
is maximally entangled, which proves the announced result.

Most of the results derived in the qubit case can be generalized
to arbitrary dimension. Let us first consider the one-repeater
configuration. Given a state $\ket{\varphi}$ (\ref{stated}), it is
always possible to transform in a deterministic way this state
into a two-qubit state of Schmidt coefficients
$(\lambda_1,1-\lambda_1)$ by local operations and classical
communication (LOCC). This follows from the application of
majorization theory to the study of LOCC transformations between
entangled states \cite{Nielsen}. Note that the SCP for the two
states is the same, $p\,^{ok}=\min(2(1-\lambda_1),1)$.

In the case of arbitrary $N$, an exponential decay for the qubit
concurrence can be shown for protocols with one-way communication.
Given an arbitrary chain, we consider the almost identical chain
where the first state is replaced by a two-qubit entangled state.
It is relatively easy to prove that the SCP decays exponentially
in the first chain if, and only if, it does it in the second one.
We start with the simplest one-repeater configuration. The
quantity to be optimized reads
\begin{equation}\label{c11}
    C_1 = 2|\det(\varphi_1)|\sup_{{\cal M}}\sum_r 2|\det(M_r \varphi_2)|,
\end{equation}
where, as above, $\varphi_1$ ($\varphi_2$) is the $2\times 2$
($d\times d$) matrix corresponding to $\ket{\varphi_1}$
($\ket{\varphi_2}$), while $M_r$ is a $2\times d$ matrix
associated to the measurement outcome $r$ at the repeater. Thus,
we recognize in the r.h.s. of Eq. (\ref{c11}) the optimal average
concurrence we can obtain out of $\ket{\varphi_2}$ by measurements
on one particle which correspond to operators of rank 2. We denote
this quantity by $\bar C$, by $\lambda_1\ge \lambda_2\ge \ldots
\ge \lambda_d$ the Schmidt coefficients corresponding to
$\varphi_2$ and by $p\,^{ok}$ its SCP, as above. For the outcome
$r$, which occurs with probability $p_r$, $\mu^r_1 \ge \mu^r_2$
denote the Schmidt coefficients corresponding to the resulting
two-qubit state $\ket{\varphi_r}$. With this notation, we have
\begin{equation}
     \bar C = 2\sum_r p_r \sqrt{\mu^r_1 \mu^r_2} \le 2\sqrt{x}\sqrt{1-x},
\end{equation}
where
\begin{equation}
     x=\sum_r p_i \mu^r_2 \le 1-\lambda_1,
\end{equation}
where the last inequality follows from the majorization criterion
\cite{Nielsen}. The optimal value is obtained for $x=p\,^{ok}/2$,
which is achieved when $p_1=1$. Thus, we obtain
\begin{equation}
  C_1 = 2|\det(\varphi_1)|\sqrt{p\,^{ok}(2-p\,^{ok})} .
\end{equation}
Note that $\sqrt{p\,^{ok}(2-p\,^{ok})}\leq 1$, with equality if
and only if $p\,^{ok}=1$. Note also, and this is important for
what follows, that the optimal strategy only depends on
$\ket{\varphi_2}$, and not on the first two-qubit state,
$\ket{\varphi_1}$.

This strategy can be generalized to the case of $N$ repeaters when
the measurements proceed from left to right. We show this
generalization for the case $N=2$, the case of arbitrary $N$ will
immediately follow. Consider the measurement step in the second
repeater. After receiving the information about the measurement
result in the first repeater $r_1$, $R_2$ has to measure his
particles. For each value of $r_1$, and since $A$ is a qubit, $A$
and $R_2$ share a two-qubit pure state, $\ket{\varphi_{r_1}}$ .
Therefore, for each measurement result, $R_2$ is back at the
previous one-repeater situation. The optimal measurement strategy
in this case was independent of the entanglement of the first
two-qubit state. Thus, up to local unitary transformations, the
measurement to be applied in the second repeater is independent of
$r_1$, and
\begin{equation}\label{c12}
    C_{2}=C_{1}|\sqrt{p\,^{ok}_3(2-p\,^{ok}_3)}| ,
\end{equation}
where $p\,^{ok}_k$ is defined as above for the state
$\ket{\varphi_k}$. It is straightforward that this reasoning
generalizes to an arbitrary number of repeaters, so
\begin{equation}\label{c1N}
    C_{N}=2|\det(\varphi_1)|\prod_{j=2}^{N+1}\sqrt{p\,^{ok}_j(2-p\,^{ok}_j)} .
\end{equation}
Therefore, the average concurrence decreases exponentially with
the number of repeaters unless the connecting pure states have
$p\,^{ok}=1$. A non-exponential decay of the SCP when $p\,^{ok}<1$
would contradict this result.

\begin{table}
  \centering
  \begin{tabular}{|c|c|}
    \hline
    Lattice & Percolation Threshold Probability \\
    \hline
    Square & $\frac{1}{2}$ \\
    Triangular & $2\sin\left(\frac{\pi}{18}\right)\approx 0.3473$ \\
    Honeycomb & $1-2\sin\left(\frac{\pi}{18}\right)\approx 0.6527$ \\
    \hline
   \end{tabular}
   \caption{Bond Percolation Threshold Probabilities
   for some examples of 2D lattices.}\label{percpr}
\end{table}

\begin{figure}
\begin{center}
  \includegraphics[width=8cm]{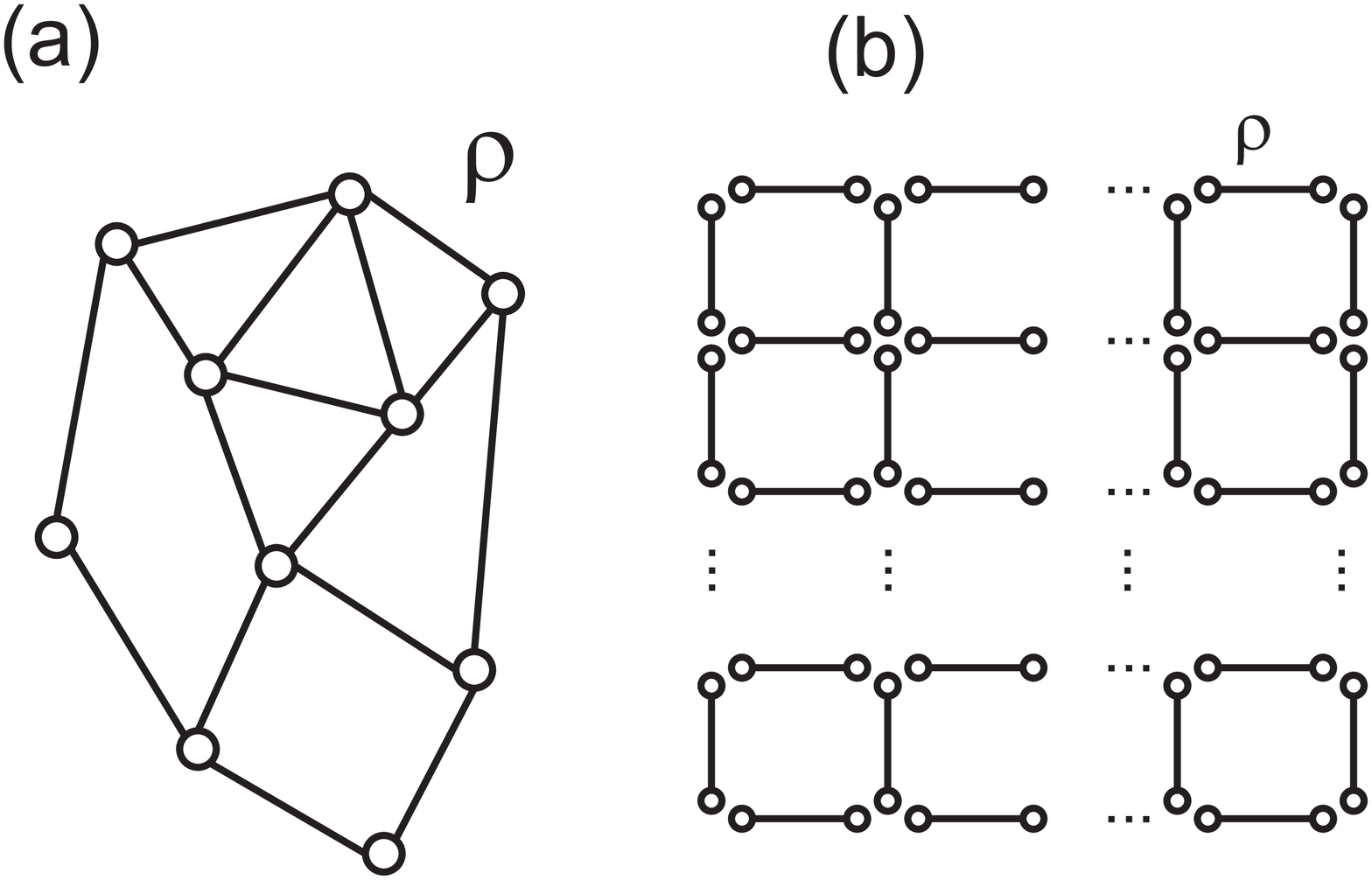}\\
  \caption{Quantum networks. A general quantum communication network consists of an arbitrary number of nodes in a given geometry sharing some quantum
  correlations, given by a global state $\rho$, as in (a). Here we consider a simplified network where the nodes are disposed according to a well-defined geometry, e.g. a 2D square lattice in (b), and where each pair of nodes is connected by the same pure state $\ket{\varphi}$.}
\label{repeater}
\end{center}
\end{figure}

\begin{figure}
\begin{center}
  \includegraphics[width=8cm]{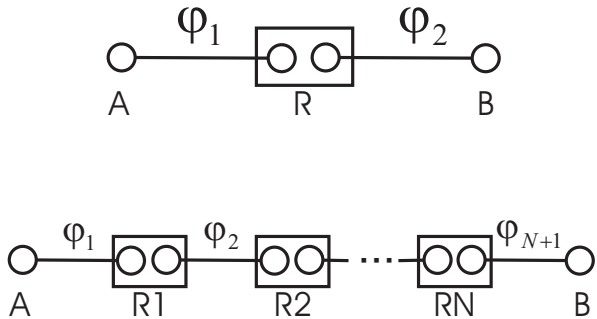}\\
  \caption{Quantum repeaters in the $1D$ chain. The upper
  figure shows the simplest one-repeater configuration, that is
  generalized below. The first step of the optimal strategy for the one-repeater configuration and qubits, where $\ket{\varphi_1}=\ket{\varphi_2}=\sqrt{\lambda_1}\ket{00}+
\sqrt{\lambda_2}\ket{11}$, consists of entanglement swapping at the repeater. The resulting states between nodes $A$ and $B$ are
$(\lambda_1\ket{00}\pm  \lambda_2\ket{11})/\sqrt{\lambda^2_1+\lambda^2_2}$ with probability $(\lambda^2_1+\lambda^2_2)/2$ and $(\ket{01}\pm\ket{10})/\sqrt 2$ with probability $p=\lambda_1\lambda_2$. Collecting all these terms, the average SCP between $A$ and $B$
is equal to $2(\lambda_1\lambda_2+\lambda_2^2)=2\lambda_2=p\,^{ok}$, which is known to be optimal. }
\label{repeater}
\end{center}
\end{figure}

\begin{figure}
\begin{center}
  \includegraphics[width=8cm]{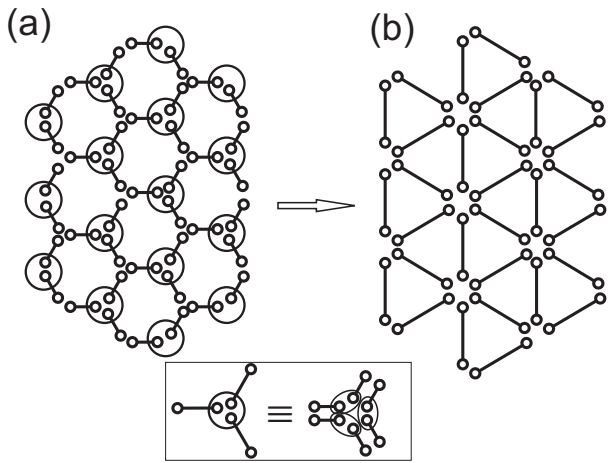}\\
  \caption{Example of quantum network where entanglement percolation and CEP are not equivalent. Each node is connected by a state consisting of two copies of the same two-qubit state, $\ket{\varphi}=\ket{\varphi_2}^{\otimes 2}$. The nodes marked in (a) perform the optimal measurement for the one-repeater configuration on pairs of qubits belonging to different connections, as shown in the inset. A triangular lattice is then obtained where the SCP for each connection is the same as for the two-qubit state $\ket{\varphi_2}$. The remaining nodes perform CEP on the new lattice.}
\label{hontr}
\end{center}
\end{figure}

\noindent{\bf Acknowledgements.} We thank  F. Verstraete, J. Wehr
and M.M. Wolf for discussions. We acknowledge support from
Deutsche Forschungsgemeinschaft, EU IP Programmes ``SCALA'' and ``QAP",
European Science Foundation PESC QUDEDIS, MEC (Spanish
Government) under contracts FIS 2005-04627, FIS 2004-05639, ``Ram\'on y Cajal" and Consolider QOIT.


\end{document}